\documentstyle[prl,aps,epsf]{revtex}

\begin{document}

\twocolumn[\hsize\textwidth\columnwidth\hsize\csname@twocolumnfalse\endcsname

\begin{center}
{\large   Instantons and the $\Delta I=1/2 $ Rule}

\it{ N.I. Kochelev$^{a,b,c}$  and  V. Vento$^a$}
\address
{\it
$^a$ Departament de F\'{\i}sica Te\`orica and Institut de F\'{\i}sica 
Corpuscular,\\
Universitat de Val\`encia-CSIC,
E-46100 Burjassot (Valencia), Spain \\
$^b$ Bogoliubov Laboratory of Theoretical Physics,\\
Joint Institute for Nuclear Research,
Dubna, Moscow region, 141980 Russia,\\
$^c$ Institute of Physics and Technology, Almaty, 480082, Kazakhstan}
\maketitle

\end{center}
\begin{abstract}
The instanton induced interaction leads to a significant
enhancement of the $A_0$ weak amplitude determining the $\Delta I=1/2$ rule,
through the contribution of operators with 
dimension $d=9$, as we show in the weak $K\rightarrow \pi\pi$ decay. 
\end{abstract}
\vspace{0.5cm}
\begin{center}
Pacs numbers: 11.30.Hv, 12.15.-y, 12.55.Ji, 12.38.Lg, 13.25.Es
\end{center}
\vspace{1cm}
]

Recent experiments confirmed the large CP violation in $K\rightarrow \pi\pi$
decays \cite{KTeV,NA48}. One of the cornerstones of this problem is famous $
\Delta I=1/2$ rule \cite{buras,reviews}. This phenomenological rule is
related to the observation of a large enhancement of the weak decays with an
isospin change of $\Delta I=1/2$ with respect to those decays with one of $
\Delta I=3/2$. Several contributions have been considered responsible for
the enhancement\cite{bertolini,cheng,another}. One of them is the well known
perturbative QCD contribution due to the exchange of hard gluons \cite{pQCD}
. It arises from short distances, and large quark and gluon virtualities.
Typically the enhancement factor of these calculations is four, far away
from the data \cite{dgh}. Another possible source of the rule comes from
long distance hadronic final state interaction (FSI) and can lead to an
enhancement of the $A_0$ amplitude which reaches about half of the
experimental value \cite{FSI,critic}.

We propose a new large contribution to the weak amplitudes arising from QCD
through the nonperturbative multi-quark 't Hooft interaction \cite{thooft1},
induced by strong fluctuations of the gluon fields known as instantons,
which strongly favors the rule. This interaction has flavor properties, very
distinct from those of the perturbative gluon exchange, which magnify the
interaction in channels , like the $I=0$ channel of the weak decays, with
vacuum quantum numbers. These properties are instrumental in the resolution
of the $U(1)_{A}$ problem \cite{thooft} explaining the large mass of $\eta
^{\prime }$ meson. We show here that the same mechanism is relevant for
understanding the $\Delta I=1/2$ rule.

A multi-quark interaction arises from the existence of quark zero modes in
the instanton field. For $N_f=3$ and for zero current quark masses,
$m_u=m_d=m_s=0$, this interaction is given by\cite{SVZ,NVZ}: 
\begin{eqnarray}
{\cal H}_{^{\prime}tHooft}& =& \int d\rho n(\rho)(4\pi^2\rho^3)^3 \frac{1}
{6N_c(N_c^2-1)}\epsilon_{f_1f_2f_3}\epsilon_{g_1g_2g_3}  \nonumber \\
& &\{\frac{2N_c+1}{2N_c+4}\bar q_{R}^{f_1}q_{L}^{g_1} \bar
q_{R}^{f_2}q_{L}^{g_2}\bar q_{R}^{f_3}q_{L}^{g_3}  \nonumber \\
&+& \frac{3}{8(N_c+2)} \bar q_{R}^{f_1}q_{L}^{g_2}\bar q_R^{f_2}
\sigma_{\mu\nu} \bar q_L^{g_2}\bar q_R^{f_3}\sigma_{\mu\nu}q_L^{g_3}
\nonumber \\
&+& (R\leftrightarrow L)\},  \label{thooft}
\end{eqnarray}
where $\rho$ is the instanton size and $n(\rho)$ is the density of the
instantons. In calculations we use the instanton liquid model for the QCD vacuum \cite
{shuryak,dp}(see reviews ref.\cite{reviews1}). For quarks with nonzero virtualities, $k_i^2$, the vertex (\ref
{thooft}) should be multiplied by the product of Fourier transformed quark
zero modes in the instanton field 
\begin{equation}
Z=\prod_i F(k_i^2),
\end{equation}
which in the singular gauge has the form 
\begin{equation}
F(k_i^2)=-x\frac{d}{dx}\left.\{I_0(x)K_0(x)-I_1(x)K_1(x)\right.\},
\label{form}
\end{equation}
where $x=\rho\sqrt{k_i^2}/2$ \cite{carlitz}.

This interaction has a large quark helicity flip $\Delta Q=2N_f$, which
comes from the definite helicity of the quarks on zero modes. Moreover the
Pauli Principle of these quarks implies that the interaction is
antisymmetric under permutations of any incoming and any outgoing quark.
This property leads to a single instanton contribution to the weak $\Delta
I=1/2$ amplitude (see Fig.1).

The standard $\Delta S=1 $ weak effective Hamiltonian is given by 
\begin{equation}
{\cal H}_{eff}^{\Delta S=1}=\sqrt{2} G_F V_{ud}V_{us}^*\sum_{i=1}^{8}
C_i(\mu)Q_i(\mu),  \label{standard}
\end{equation}
where the $Q_i$ operators, with dimension $d=6$, used are those of \cite
{pQCD} and the coefficients $C_i(\mu)$ come from calculation of the hard
perturbative gluon coupling to the weak amplitudes. The scale $\mu\approx 1
GeV$ of these calculations determines the kinematical region where one can
believe pQCD, and the lower limit of the integration over the virtuality of
the quarks and gluons in loops created by gluon exchanges. The matrix
elements of the $Q_i(\mu)$ operators in (\ref{standard}) are calculated
within nonperturbative approaches, for example by bosonising the quark
operators or by performing lattice simulations.

\begin{figure}
\epsfxsize=9cm
\epsfysize=3cm
\centerline{\epsfbox{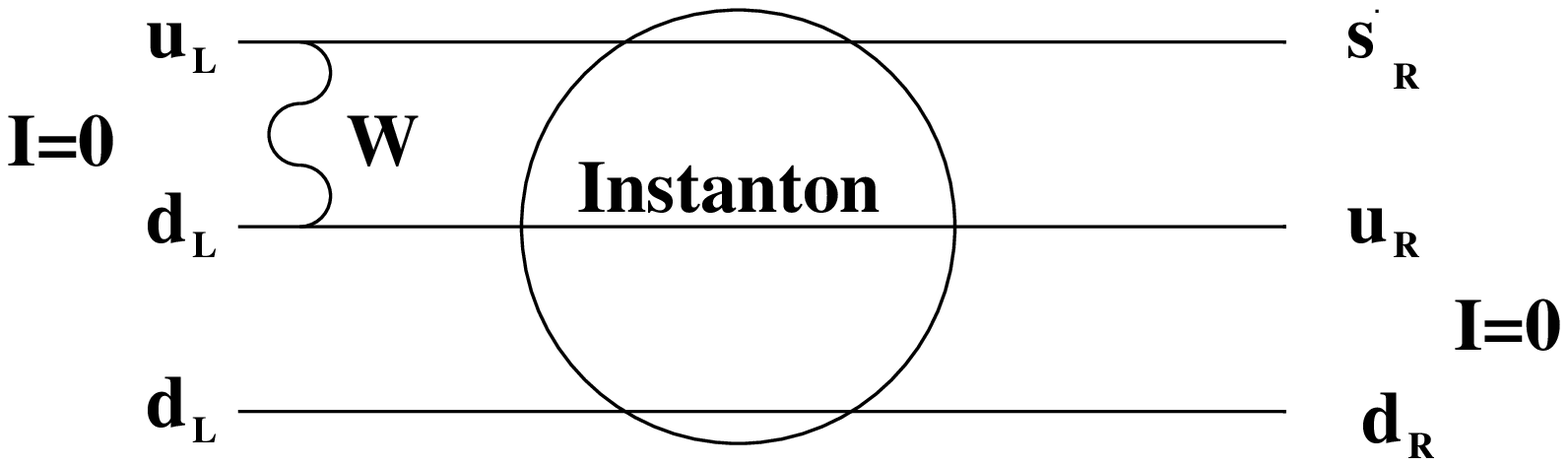}}
\end{figure}
Fig.1. The contribution of the six-quark instanton induced interaction to
the $\Delta I=1/2$ weak amplitude. $W$ denotes the W-boson exchange.\newline

Our important observation is that there is some additional term in the weak
interaction Hamiltonian with $\Delta S=1 $, coming from the six-quark
interaction (Fig.1), which corresponds to the operator of dimension $d=9$, 
\begin{eqnarray}
{Q}_{I+\bar I}^{d=9}&=& \frac{2N_c+1}{2N_c+4}\bar u_{R}d_{L} \bar s_{R}u_{L}
\bar d_{R}d_L  \nonumber \\
&+& \frac{3}{8(N_c+2)} \bar u_{R}d_{L}\bar s_R\sigma_{\mu\nu} \bar u_L\bar d
_R\sigma_{\mu\nu}d_L  \nonumber \\
&+& (-1)^P perm.(u_R,d_R,s_R) +(R\leftrightarrow L)  \label{thoofto}
\end{eqnarray}
where $P$ is number of the quark permutations and which contributes only to
$\Delta I=1/2$ transitions. Compared with the gluon induced operators (\ref
{standard}), this operator violates helicity conservation.

With respect to the scale of the new six-quark operators a comment is
needed. We would like to treat the operators in (\ref{standard}) as local,
therefore the integration over the quark virtualities in the loop of Fig.1
should be limited by the hadronization scale $\tilde{\mu}\approx
\Lambda_{QCD}\approx 1/R\approx m^*=260 MeV$, where $m^*=-2\pi^2\rho_c^2
<0|\bar qq|0>/3$ is the effective quark mass in the instanton vacuum.

One of the manifestations of the $\Delta I=1/2$ rule is a huge enhancement
of the $K\rightarrow \pi\pi $ amplitude in the isospin- zero state $A_0$ as
compared with amplitude to the isospin-two $\pi\pi$, $A_2$, i.e., $22.2$.

Our instanton induced weak interaction, which only contibutes to the $A_0$
amplitude, is shown in Fig.2, for the $K^0$ decays. We use the normalization
for the $K\rightarrow\pi\pi$ amplitude of Bel'kov et al. in \cite{reviews} 
\begin{eqnarray}
M_{K^0\rightarrow\pi^+\pi^-} &=&\sqrt{\frac{2}{3}}A_0e^{i\delta_0}+ \frac{1}
{\sqrt{3}}A_2e^{i\delta_2}  \nonumber \\
M_{K^0\rightarrow\pi^0\pi^0}&=&\sqrt{\frac{2}{3}}A_0e^{i\delta_0}- \frac{2}
{\sqrt{3}}A_2e^{i\delta_2}  \nonumber \\
M_{K^+\rightarrow\pi^+\pi^0}&=& \frac{\sqrt{3}}{2}A_2e^{i\delta_2}.
\label{ampli}
\end{eqnarray}
\begin{figure}
\epsfxsize=7.5cm
\epsfysize=3cm
\centerline{\epsfbox{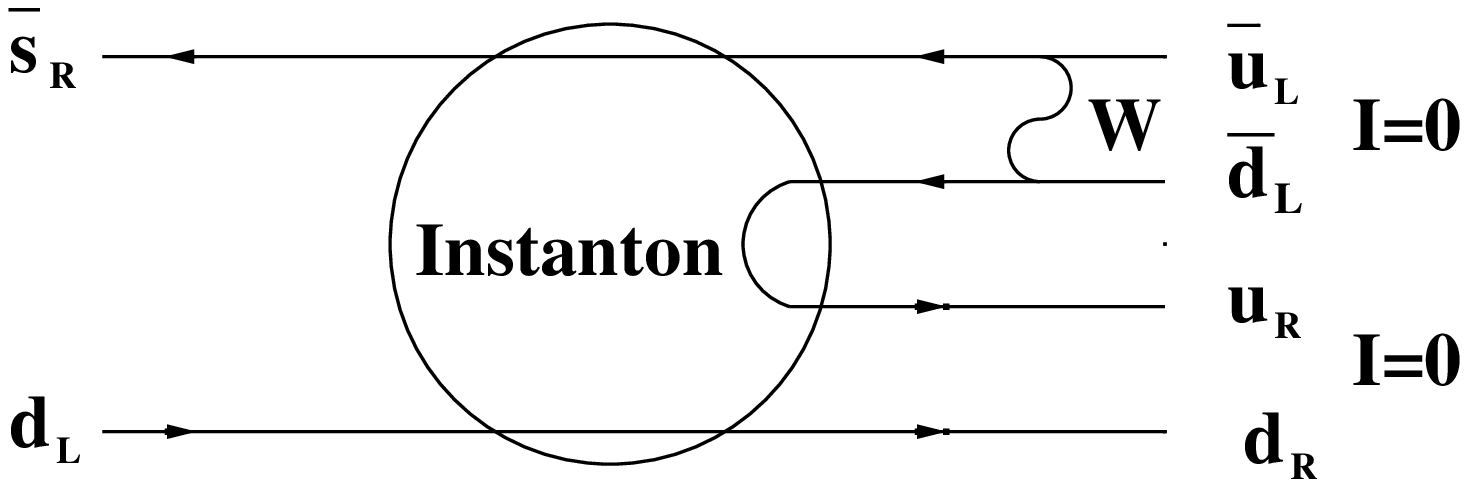}}
\end{figure}
Fig.2. The contribution of the six-quark instanton induced interaction to
the $K^0\rightarrow\pi\pi$ decay.\newline

The calculation of the diagram in Fig.2 gives for the induced effective
Hamiltonian, responsible for the $K^0\rightarrow\pi^+\pi^-$ decay and $N_c=3$
, 
\begin{eqnarray}
{\cal H}_{inst}^{K^0\rightarrow\pi^+\pi^-}& =& \frac{C_1(\mu)G_F}{\sqrt{2}}
V_{ud}V_{us}^*\int d\rho \frac{n(\rho)}{2\pi^2} (\frac{4\pi^2\rho^3} {3})^3 
\nonumber \\
& &\int_{\tilde{\mu}}^\mu dkkF^2(k\rho/2) \bar u_Rd_L\{\bar s_{R}u_{L} \bar d
_{R}d_{L}  \nonumber \\
&+& \frac{3}{32}(\bar s_{R}\lambda^au_{L} \bar d_{R}\lambda^ad_{L}  \nonumber
\\
&-&\frac{3}{4} \bar s_R\sigma_{\mu\nu}\lambda^a u_L\bar d_R\sigma_{\mu\nu}
\lambda^ad_L)\},  \label{K0}
\end{eqnarray}
where $C_1(\mu)$ is the Wilson coefficient of $Q_1$ with $I=0$ in (\ref
{standard}), which is related to the pQCD contribution for values of the
quark virtualities between $\mu$ and $M_W$.

By using appropriate Fierz transformations and making the PCAC substitutions 
\begin{eqnarray}
\bar d\gamma_5 u&=&\frac{-i\sqrt{2}F_\pi m_\pi^2}{m_u+m_d}\phi_{\pi^+}, 
\nonumber \\
\bar u\gamma_5 d&=&\frac{-i\sqrt{2}F_\pi m_\pi^2}{m_u+m_d}\phi_{\pi^-}, 
\nonumber \\
\bar s\gamma_5 d&=&\frac{-i\sqrt{2}F_K m_K^2}{m_s+m_d}\phi_{K^0},
\label{subs}
\end{eqnarray}
where $F_\pi=93 MeV$, we arrive at the following matrix element 
\begin{eqnarray}
M_{K^0\rightarrow\pi^+\pi^-}&=& -C_1(\mu){G_F}V_{ud}V_{us}^*\frac{11}{96\pi^2
}  \nonumber \\
& &(\frac{F_\pi m_\pi^2}{m_u+m_d})^2 (\frac{F_Km_K^2}{m_s+m_d})  \nonumber \\
& &\int d\rho n(\rho)(\frac{4\pi^2\rho^3}{3})^3\int_{\tilde{\mu}}^\mu
dkkF^2(k\rho/2).  \label{answer}
\end{eqnarray}

We obtain an estimate of the instanton contribution by using Shuryak's 
instanton liquid model \cite{shuryak} with the density given
by 
\begin{equation}
n(\rho)=\frac{n_{eff}}{(m^*\rho)^3}\delta(\rho-\rho_c)  \label{dens}
\end{equation}
and 
\begin{equation}
n_{eff}=1fm^{-4}, {\ }\rho_c=1.6 GeV^{-1}  \label{param}
\end{equation}

In chiral limit, $m_{u}=m_{d}=m_{s}=0$, from (\ref{answer}) we obtain 
\begin{eqnarray}
M_{K^{0}\rightarrow \pi ^{+}\pi ^{-}} &=&-C_{1}(\mu ){G_{F}}
V_{ud}V_{us}^{\ast }\frac{11}{12\pi ^{2}F_{\pi }^{3}}n_{eff}  \nonumber \\
&&\int_{\tilde{\mu}}^{\mu }dkkF^{2}(k\rho _{c}/2)  \label{final}
\end{eqnarray}
where the Gell-Mann-Oakes-Renner relations 
\begin{eqnarray}
F_{\pi }^{2}m_{\pi }^{2} &=&-m_{u}<0|\bar{u}u|0>-m_{d}<0|\bar{d}d|0>, 
\nonumber \\
F_{K}^{2}m_{K}^{2} &=&-m_{s}<0|\bar{s}s|0>-m_{d}<0|\bar{d}d|0>  \label{rel}
\end{eqnarray}
have been used.

Therefore our final result for the six-quark instanton interaction
contribution to $A_0$ amplitude is 
\begin{eqnarray}
A_0^{d=9}&=& -\frac{\sqrt{3}C_1(\mu)G_F}{\sqrt{2}}V_{ud}V_{us}^*\frac{11}{
12\pi^2F_\pi^3} n_{eff}  \nonumber \\
& &\int_{\tilde{\mu}}^\mu dkkF^2(k\rho_c/2),  \label{final2}
\end{eqnarray}

With the chosen value of the parameters (\ref{param}) and LO value $
C_1(1GeV)\approx c_1(1GeV)-c_2(1GeV) \approx 1.9 $, where the values of $c_1$
and $c_2$ are given in \cite{reviews} for $\Lambda^{(4)}_{\overline{MS}}=215
MeV $, we have for ratio 
\begin{equation}
\frac{A_0^{d=9}}{A_0^{exp}}= 0.5,  \label{ratio}
\end{equation}
where the $A_0^{exp}$ is the experimental amplitude \cite{FSI1}.

The contribution of the instantons leads to a strong enhancement of the $
A_{0}$ amplitude in weak $K$ meson decays. Let us discuss the various
contributions to the final number by using the, large $N_{C}$, $A_{0}$
amplitude 
\begin{equation}
A_{0}^{N_{c}\rightarrow \infty }=-\sqrt{\frac{3}{2}}G_{F}V_{ud}V_{us}^{\ast
}F_{\pi }(m_{K}^{2}-m_{\pi }^{2})
\end{equation}
as a scale. The pQCD corrections provide us with a factor of about 1.9,
while the pure instanton contribution leads to a factor of about 2.0. One
should not forget that additional contributions to the ratio, for example
FSI, will further increase it.

Away from chiral limit there are corrections to Eq.(\ref{ratio}) 
coming from additional  terms proportional to the current quark masses 
in the PCAC relations Eq.(\ref{subs}). Moreover there are other 
contributions arising also from $SU(3)$ breaking in $K\rightarrow\pi\pi$ 
amplitude, e.g., those arising from $d=6$ instanton induced
operators, others from the quark non-zero mode
contributions to the instanton field, etc. Their analysis is beyond 
the skope of this paper. Anyway we expect the chiral expansion to be 
relatively soft and the ratio to change  at most at the
level of  $m_K^2/\Lambda_\chi^2 \approx 25\%$.

The most intriguing term, because of its lower
dimensionality, is the one with dimension $d=6$. Its 
corresponding operators  arise from the reduction of the six-quark 
't Hooft interaction  to  a four-quark interaction by closing  one 
of the quark lines  by a  quark condensate. In the $SU(3)_f$  limit 
and for $N_c=3$  the effective $\Delta S=1$ lagrangian term induced 
by  this interaction has the  form Eq.(\ref{standard}) with
\begin{eqnarray}
{Q}_{I+\bar I}^{d=6}&=& \bar u_{R}d_{L} \bar s_{R}u_{L}
+\frac{1}{4} \bar u_{R}\sigma_{\mu\nu}d_{L}\bar
s_R\sigma_{\mu\nu}u_L
\nonumber \\
& &-(u_R\leftrightarrow d_R) +(R\leftrightarrow L) 
\label{thoofd6}
\end{eqnarray}
and
\begin{eqnarray}
C(\mu)^{d=6}&=& 
\frac{2C_1(\mu)n_{eff}}{3<0|\bar qq|0>^2\pi^2}\nonumber \\
& &\int_{\tilde{\mu}}^\mu dkkF^2(k\rho_c/2), 
\label{finald6}
  \label{Cd6}
\end{eqnarray}
The use of the vacuum insertion method \cite{insertion} shows that 
in chiral limit the contribution of this operator  to 
$K^0\rightarrow\pi^+\pi^-$  decay amplitude is zero.

We have shown that a novel mechanism arising from the $N_{f}=3$ 
't Hooft instanton induced interaction contributes considerably to the
empirical $\Delta I=1/2$ rule found in the weak $\Delta S=1$ decays. This
instanton induced multi-quark interaction, due to its specific flavor
dependence, is able to contribute to the strong enhancement of $A_{0}$
amplitude in $K\rightarrow \pi \pi $ decays. Moreover it proclaims the
importance of the contribution of higher dimensional operators \cite{don},
in particular $d=9$ in our case, and the quantum numbers of the instanton
induced interaction, in the weak decays. 

We should mention that recently an attempt to incorporate
instanton physics into the description of weak processes
has been carried out by Franz et al.\cite{franz}
within the Chiral Quark Model of Diakonov and Petrov
\cite{dp,diakonov}.  This approach takes into account only
those terms of the $N_{f}=2$ instanton  induced lagrangian
which are leading order in the $1/N_{C}$ expansion and 
therefore does not consider the mechanism we propose here,
based  on the full $N_{f}=3$ instanton induced lagrangian. Our 
mechanism is an alternative to that 
proposed by  Franz et al. which requires, in order to produce 
a relevant ratio, anomalously small  values of the constituent 
quark masses at zero virtuality, thus destroying  the previous 
achievements of the Diakonov-Petrov scheme.

Our calculation has been performed in the chiral limit. Much work needs to
be done to relax this limit. We have discussed some of the new mechanisms
one might encounter. However, the fact that in the chiral limit our result is
quantitatively relevant signals the importance of instanton physics in this
field. One may thus conclude, that direct instanton contributions of the type
discussed here cannot be omitted in any serious study of the non-leptonic
decays and are important in closing the gap between the theoretical
interpretation and the experimental value.

We are grateful to A.E.Dorokhov, A.A.Bel'kov, F.Botella, H.-Y.Cheng,
V.Gim\'enez, T.Hambye, A.Pich and W.M.Snow for useful comments. One of us
(N.I.K) is very grateful to the University of Valencia for the warm
hospitality. This work was partially supported by
DGICYT PB97-1227.


\begin{references}
\bibitem{KTeV}  A.Alavi-Harati et al., Phys. Rev. Lett. {\bf 83}, 22 (1999).

\bibitem{NA48}  V.Fanti et al., Phys. Lett. {\bf B465}, 335 (1999);
T.Gershon, On behalf of the NA48 Collaboration, hep-ex/0101034.

\bibitem{buras}  A.J. Buras, hep-ph/0101336.

\bibitem{reviews}  G.Buchalla, A.J.Buras and M.E.Lantenbacher,\newline
Rev. Mod. Phys. {\bf B68}, 1125 (1996) ; S.Bertolini, M.Fabbrichesi and
J.O.Eeg, Rev. Mod. Phys. {\bf 72}, 65 (2000); E. de Rafael, hep-ph/9502254;
J.Bijnens, hep-ph/0010265; L.Lellouch, hep-lat/0011088; A.A.Bel'kov et al.,
hep-ph/9907335.

\bibitem{bertolini}  S.Bertolini, hep-ph/0001235.

\bibitem{cheng}  H.-Y. Cheng, Int. J. Mod. Phys. {\bf A4}, 495 (1989).

\bibitem{another}  S.Bertolini et al., Nucl.Phys. {\bf B514}, 63 (1998);
T.Hambye, G.O.Koehler and P.H. Soldan, Eur. Phys. J. {\bf C10}, 271 (1999);
Y.-L. Wu, hep-ph/0012371.

\bibitem{pQCD}  V.I.Vainstein, V.I.Zakharov and M.A.Shifman,\newline
JETP {\bf B72}, 1275 (1977); M.A.Shifman, V.I.Vainstein and V.I.Zakharov,
Nucl. Phys. {\bf B120}, 316 (1977); M.K.Gaillard and B.W. Lee, Phys. Rev.
Lett. {\bf 33}, 108 (1974); G.Altarelli and L.Maiani, Phys. Lett. {\bf 52B},
351 (1974).

\bibitem{dgh}  J.F.Donoghue, E.Golowich and B.R.Holstein, {\it Dynamics of
the Standard Model}, (Cambridge University Press, Cambridge, England 1992).

\bibitem{FSI}  E.Pallante and A.Pich, {\it Phys. Rev. Lett. } {\bf 84}, 319
(2000); Nucl. Phys. {\bf B592}, 294 (2000); E.A.Pashos, hep-ph/9912230;
T.N.Truong, hep-ph/0004185.

\bibitem{critic}  A.J.Buras et al., Phys. Lett. {\bf B480}, 80 (2000).

\bibitem{thooft1}  G.'t Hooft, Phys. Rev. {\bf D32}, 3432 (1976).

\bibitem{thooft}  G.'t Hooft, Phys. Rev. {\bf D14}, 3432 (1976).

\bibitem{SVZ}  M.A.Shifman, A.I.Vainshtein, A.I.Zakharov, Nucl. Phys.
{\bf B163}, 43 (1980).

\bibitem{NVZ}  M.A.Nowak, J.J.M.Verbaarschot and I.Zahed, Nucl. Phys.
{\bf B324}, 1 (1989).

%
\bibitem{shuryak}  E.V. Shuryak, Nucl. Phys. {\bf B203}, 93 (1982); ibid 116 ;
ibid 140; Nucl. Phys. {\bf B214}, 237 (1983).

\bibitem{dp} D. Diakonov and V. Yu. Petrov, Phys. Lett. {\bf B147}, 351 (1984);
Nucl.Phys. {\bf B245}, 259 (1984); Sov. Phys. JETP {\bf 59}, 13 (1984); 
Nucl. Phys. {\bf B272}, 457 (1986).

\bibitem{reviews1}
E.V. Shuryak, Phys. Rep. {\bf 115}, 151 (1984); T. Sch\"afer and E.V. Shuryak,
Rev. Mod. Phys. {\bf 70}, 1323 (1998).

\bibitem{carlitz} R.D. Carlitz, Phys. Rev {\bf D17}, 3225 (1978);
D. Diakonov and V.Yu. Petrov, Sov. Phys. JETP {\bf 62}, 204 (1985).
\bibitem{insertion} B.W. Lee, J.R. Primack and S.B. Treiman,
Phys. Rev. {\bf D7}, 510 (1973); M.K. Gaillard and B.W. Lee,
Phys. Rev. {\bf D10}, 897 (1974).

 
\bibitem{FSI1}  A.Pich, B. Guberina and E. de Rafael, Nucl. Phys.
{\bf B277}, 197 (1986).

\bibitem{don}  V. Cirigliano, J.F. Donoghue and E. Golowich, hep-ph/0007196.

\bibitem{franz}  M. Franz, H.-C. Kim and K. Goeke,
hep-ph/9908400; Nucl. Phys. {\bf A663-664}, 995 (2000) ;
Nucl. Phys. {\bf B562}, 213 (1999).



\bibitem{diakonov} D. Diakonov and V. Yu. Petrov,
JETP. Lett. {\bf38}, 433 (1983); 
JETP. Lett. {\bf 43}, 75 (1986); D. Diakonov,
V. Yu. Petrov and P.V. Pobylitsa, Nucl. Phys. {\bf B306}, 809 (1988).
\end{references}
\end{document}